\documentstyle[aps,epsfig,pre,twocolumn]{revtex}
\begin{document}
\draft

\title{Local minimal energy landscapes  in river networks}
\author{Achille Giacometti}

\vskip 2.0cm
\address{INFM Unit\'a di Venezia, Dipartimento
di Scienze Ambientali, Universit\'a di Venezia \\ 
Calle Larga Santa Marta DD2137, I-30123 Venezia-Italy}

\date{\today}

\maketitle

\begin{abstract}
The existence and stability of the universality class associated to local
minimal energy landscapes is investigated.
Using extensive numerical simulations, we first study
the dependence on a parameter $\gamma$ of
a partial differential equation which was proposed to describe
the evolution of a rugged landscape toward a local minimum of the dissipated 
energy.
We then compare the results with those obtained by an evolution scheme based on
a variational principle (the optimal channel networks).
It is found that both models yield qualitatively similar river patterns and
similar dependence on $\gamma$. The aggregation mechanism
is however strongly dependent on the value of $\gamma$.
A careful analysis suggests that scaling behaviors may weakly depend both on 
$\gamma$ and on initial condition, but in all cases it is within 
observational data predictions. Consequences of our results
are finally discussed and the most plausible scenario is presented.
\end{abstract}
%

\section{Introduction}
\label{sec:introduction}
Understanding the development of a landscape in terms of fundamental mechanical
principles is a formidable task \cite{Iturbe97}. 
In spite of this high complexity, recent
theoretical studies have resulted in considerable progresses by considering
the issue from a viewpoint analogous to the one taken in conventional critical
phenomena \cite{Rinaldo93,Meakin91,Maritan96_1,Banavar97,Sinclair96,Somafai97,Dodds99} 
where simple models are exploited to
identify universality classes. 
The main idea of this approach is indeed to focus on few fundamental
ingredients which, in the spirit of critical phenomena, are expected to
provide a reasonable description of the large-scales properties of
the network.

The remarkable properties of river networks have been known for some time
and they can be condensed in few phenomenological scaling laws which
have been confirmed to hold in the observational data \cite{Leopold62}.
While these laws do not explain the underlying physical mechanisms, they
nevertheless provide guidelines for their search.
Hence any newly proposed model for river networks ought to be tested against
these laws. In the language of critical phenomena, those scaling laws
can be used to derive critical exponents and thus discriminate among 
different universality classes. Among such laws a fundamental role in the
physics of the erosion of a landscape due to the flow of water over it, is
played by the slope-area law (see Ref. \cite{Iturbe97} for a review).

A few Langevin-like equations have been recently proposed for describing the 
evolution of the landscape under the effect of erosional processes.
In Ref. \cite{Banavar97}, reparametrization invariance arguments 
\cite{Marsili96} were used to derive
a dynamical equation which yields the slope-area law as a stationary state.
The same equation was obtained by Somafai and Sander \cite{Somafai97}
using Landau arguments. Other proposals have also been advanced 
\cite{Sinclair96},\cite{Pastor98,Metiver99,Giacometti95}.

The equation proposed in Ref. \cite{Banavar97}, was studied both analytically
in 1+1 dimensions (one spatial and one temporal) and numerically using a 
self-consistent (SC) solution in 2+1 dimensions,  
and it was shown to
predict a fairly reasonable stationary state quite different from the
starting network acting as initial condition.
This type of analysis hinges on the observation  that the network
representing the river has an evolution mirroring that of the 
evolution profile. Other methods have also been devised which directly
address the topological properties of the network itself, the best
known being the optimal channel networks (OCN) \cite{Iturbe92}. 
This is a lattice model where a functional describing the dissipated
energy is minimized in order to find the optimal configuration and it is
based on the idea that, presumably, the erosional process taking place
on a landscape is driven by a strive for optimality. A simulating
annealing procedure \cite{Kirkpatrick83} has been implemented to this
aim both at zero \cite{Rinaldo93} and finite temperature 
\cite{Sun94}. While the latter is 
aimed to find the absolute minimum, the former is expected to display
local minima only. Using exact bounds and finite-size scaling, Maritan et al 
(\cite{Maritan96_2,Colaiori97}) showed that the absolute minimum 
belongs to the mean-field universality class, that, in turn, means that the 
corresponding network
has a highly symmetric pattern with  small rivers draining into bigger rivers
in a predictable way (this network is akin to the Peano basin 
\cite{Mandelbrot83}). However this minimum is not easily reachable in the
space of all configurations, and one is then led to suspect
that real rivers are better described by configurations related to
local minima.
We further note that stationary states of the aforementioned dynamical 
equations, are also expected to be associated with local minima 
for the same reason.

In view of the above discussion, it is apparent that a more 
complete description
of the stability and the scaling behavior associated with these
local minima would be desirable. The present
work is an attempt in this direction. Our aims are threefold. Our first goal is
the full characterization of the possible universality class associated
with local minima. Using the SC numerical procedure envisaged
in Ref. \cite{Banavar97}, we extend previous numerical results both in
size and statistics and we use a finite size procedure for a more accurate
estimate of the exponents. We then compare these exponents with 
those calculated
using a zero-temperature simulation in the OCN framework with
similar sizes and statistics.
Our second goal is 
the generalization of the dynamical equation to include a tunable parameter
which turns out to be related to the $0\le \gamma \le 1$ parameter considered 
in OCNs and which is responsible of the aggregation mechanism. In Ref.
\cite{Colaiori97} it was shown that the {\it absolute} minimum is insensitive
to a variation of $\gamma$ in the interval $1/2 \le \gamma < 1$ \cite{note1}.
We find that although the final patterns display marked differences
as a function of $\gamma$, critical exponents show a much smaller
discrepancy,  which our results indicate
to be marginal (at the edge of error bars), both in the SC and OCN case.
Finally we test the stability of our results against a 
variation in the initial 
conditions. We find that even in this case a marginal difference
appears in the effective critical exponents. 

The paper is organized as follows. In Sec.\ref{sec:DEQ}, we review the
dynamical equation whereas in Sec.\ref{sec:Exp} scaling laws and
critical exponents are briefly recalled. Sec.\ref{sec:OCN} contains the
definition of OCN and our results are presented in Sec.\ref{sec:Results}.
Finally Sec.\ref{sec:conclusions} contains some concluding remarks and future
perspectives.

 
\section{The dynamical equation }
\label{sec:DEQ}
A rather general dynamical equation consistent with
general principles, and capturing the physics of erosional processes occurring
in river basin, is  \cite{Banavar97},\cite{Somafai97}:
\begin{eqnarray}
\frac{\partial h(x,t)}{\partial t}&=& u+ \nu \nabla^2 h(x,t)
-Q^{\alpha}(x,t) \Bigl[ \nabla h(x,t)\Bigr]^2+\eta(x,t)
\label{sc1}
\end{eqnarray}
Here $h(x,t)$ and $Q(x,t)$ are the height of the landscape 
and the flow rate at position $x$ and time $t$, $\eta(x,t)$ is a noise term and
$\alpha$ is a parameter which will be related to $\gamma$ (see below).
The constant term $u$ 
on the right-hand-side of Eq. (\ref{sc1}) mimics the so-called geological
uplift which is known to originate by tectonic forces. The second term
represents a local diffusion term of strength $\nu$, condensing both smoothing 
(rounding of hilltops) and sedimentation (filling of valleys) processes.
The third term is non-local and corresponds to an erosion driven by the 
waterflow on a surface with $Q(x,t)$ representing the flow through site $x$.
The exponent $\alpha$ is a parameter of the model which was
assumed to be equal to $1$ in Ref.\cite{Banavar97} and Ref.\cite{Somafai97}.
The last noise term is added to account for small-scale stochastic processes
such as rainfall fluctuations. In the attempt
of extracting the basic features associated with the third term, one can ignore
both the diffusion and noise terms (the accuracy of this
approximation has been discussed in Ref.\cite{Banavar97} ).
In this case, a stationary
state is obtained when the uplift term balances the flow dependent
erosional term, that is when
\begin{eqnarray}
|\nabla h(x,t)| &\sim& Q^{\gamma-1}(x,t)
\label{sc2}
\end{eqnarray}
where, for later convenience, we define $\gamma=1-\alpha/2$.
Most of previous work was performed with $\alpha=1$ ($\gamma=1/2$) 
(there are few exceptions but with aims and methodologies different
from ours, see e.g. Ref. \cite{Iturbe97}).
In this case Eq. (\ref{sc2}) is known as slope-area law \cite{Iturbe97}, 
which is a robust empirical
law verified by real field observational data. We shall see later on
how $\alpha$ is related to the dissipated
energy functional. Here we note that if we assume (as we shall do
hereafter) uniform rainfall
(and no ground water), then $Q(x,t) \sim a(x,t)$ where $a(x,t)$ is the
area of the basin draining into point $x$ at time $t$. 
The basic result of this dynamical equation is that the non-linear
term is able to account for the correct relationship between local
slope and water flowrate.

\section{Critical exponents and scaling laws}
\label{sec:Exp}
River networks are a remarkable example of systems governed by scaling laws.
Although the concept of power and scaling laws has been known to the
hydrologists for half a century, it was not until recently that this concept
was put into a well defined framework \cite{Maritan96_1},\cite{Colaiori97}
in analogy with conventional non-equilibrium critical
phenomena where power laws are associated with critical exponents. For the
sake of completeness, we shall briefly review here a few of the central
laws appearing in river networks which we regards as the most fundamental.

Hack observed that the total area $a$ draining into a given point
and the upstream length $l$ going from that point to the source
though the path of maximum water flow, were not independent
but related by \cite{Hack57}
\begin{eqnarray}
l \sim a^h
\label{cr1}
\end{eqnarray}
where $h$ is often referred as Hack's exponent and it ranges in the
interval $[0.5-0.6]$ in real rivers. The distributions of drainage
areas and upstream lengths also follow a power law. Within the context
of finite-size scaling, these may be written as 
\cite{Meakin91,Maritan96_1}
\begin{eqnarray}
p(a,L)&=& a^{-\tau} f(a/L^{\phi})
\label{cr2}
\end{eqnarray}
for areas and
\begin{eqnarray}
\pi(l,L)&=& l^{-\psi} g(l/L^{d_l})
\label{cr3}
\end{eqnarray}
for lengths.
In Eq. (\ref{cr2}) $p(a,L)$ is the distribution of drainage areas
on a basin of size $L$ and $f(x)$ is a finite size function. It defines
two critical exponents $\tau$ and $\phi$ which are not independent
\cite{Maritan96_1}. Similarly $\pi(l,L)$ is the distribution
of the upstream lengths on a basin of size $L$, $g(x)$ is a finite
size function and there exists a relation between $\psi$ and $d_l$. Note
that $\phi$ and $d_l$ define the ``typical'' area 
($a_{\text{typ}} \sim L^\phi$) and length ($l_{\text{typ}} \sim L^{d_l}$).
Finally we remark that one usually distinguishes between self-affine
basins ($d_l=1$, $\phi<2$) and self-similar basins ($d_l>1$, $\phi=2$)
\cite{Cieplak98} and in both cases only one exponent out of four
is independent. They are related via scaling relations as \cite{Maritan96_1},
\cite{Colaiori97},\cite{Cieplak98}
\begin{eqnarray}
h=\frac{d_l}{\phi} \qquad \tau=2-h \qquad \psi=\frac{1}{h}
\label{cr4}
\end{eqnarray}


\section{Zero-temperature OCN: Local minima }
\label{sec:OCN}
Within the context of the OCNs, 
the ``dissipated energy'' of a river network at a given time,
can be written, in continuum notations, as \cite{Iturbe92}:
\begin{eqnarray}
E(t)&=& \int~dx~ |\nabla h(x,t)| Q(x,t)
\label{ocn1}
\end{eqnarray}
The local gradient $|\nabla h(x,t)|$ and the
water flow rate $Q(x,t)$ are expected to satisfy Eq. (\ref{sc2}). Hence
one gets:
\begin{eqnarray}
E(t)&=& \int~dx~  Q^{\gamma}(x,t)
\label{ocn2}
\end{eqnarray}
We note that although the energy expression (\ref{ocn2}) 
has been derived in the context of OCN \cite{Rinaldo93}, in principle
it could be defined in the 
evolution equation (\ref{sc1}) as well, thus providing a way of
monitoring the rate of approach to steady state.

The basic assumption of the OCN is that there is a tendency of any real
river basin, to assume a configuration which minimizes
Eq. (\ref{ocn2}). A natural question arising is the characterization
of the local and absolute minima of $E$.
It was shown \cite{Maritan96_2,Colaiori97}, that the 
{\it absolute minimum} of $E$ is insensitive to a variation of the value
of $\gamma$ in the interval $1/2 \le \gamma < 1$ \cite{note1} and then
$h=1/2$, $\tau=3/2$ and $d_l=1$.
The analogous issue in the case of local minima is much less clear.
In fact, when $\gamma=0.5$, numerical work \cite{Colaiori97} suggests
that there exists a set of local minima which presumably corresponds
to a new universality class that is the relevant one for real rivers.
Indeed only exceptional events are able to radically modify river courses
and so local minima of dissipated energy 
trap the system  with high probability and therefore dominate the statistics.

It is then a vital issue to discriminate 
whether or not local minima configurations
are indeed related to a well defined and robust universality class independent
on $\gamma$.


\section{Results}
\label{sec:Results}
In this section, we describe numerical procedures and results
in detail for each case. All calculations are carried out on
a $L \times L$ square lattice with periodic boundary conditions
in one direction which we identify as the
transverse direction. Multiple outlets are allowed in the
outflowing longitudinal direction (West side in the figures) 
whereas an infinite wall is
set up on the opposite side (East side). All averages considered in the
statistics are carried out only over the river with the largest flow.
It is worth stressing that the choice of considering only
the maximum river in a multiple outlets environment
corresponds to consider the statistical behavior of rivers that
are in competition to drain a given region. On the other hand, a single river 
within a given region is more appropriate if geological constraints
are known to exist.

Typically 8 nearest-neighbors (nn) - 4 associated with the square lattice
and 4 associated with the two diagonal directions - are allowed.
A somewhat more restricted choice considers only the 4 natural
nn associated with the square lattice structure. In view of universality
one would expect that the details of the lattice structure should not
matter. This second choice has the considerable advantage of being 
less time consuming for numerical purposes. For this reason, although
all following figures display patterns obtained with 8 nn,
the results reported in this work are obtained with statistics based on 
4 nn. In one example, we have
explicitly checked that the outcomes using the two
choices are consistent within the statistical errors.

Finally, in all our networks, the drainage area $a(x,t)$ 
is computed, at each time step, in a
standard way according to \cite{Iturbe97}:
\begin{eqnarray}
a(x,t)&=& \sum_{y(x)} a(y,t) +1
\label{int1}
\end{eqnarray}
where $y(x)$ denotes all sites $y$ which drain into $x$ and the last 
term on the
right hand side represents a unit rainfall input on each site at all times.

\subsection{Initial condition}
\label{subsec:IC}
There are many possible initial conditions that can be used in 
numerical analysis of river networks. A popular one among 
hydrologists is a deterministic comb-like structure \cite{Iturbe97}.
However this initial condition suffers of various drawbacks \cite{Caldarelli97}
and its use in the multiple outlets case, such the one treated
here, appears to be somewhat inconvenient.
Hence we consider here two other physically reasonable choices,
namely a spanning tree
(ST) and a Scheidegger network (SN). Although STs are well known
in statistical physics mainly due to their relation with the $q \to 0$ limit
of the Potts model \cite{Wu82}, only recently 
their topological properties have been studied in details \cite{Manna92}. A suitable
variation of spanning trees has even been proposed as a topological model for
river networks \cite{Cieplak98},\cite{Manna96}. We have generated
STs with multiple outlets by using an adapted Broder's algorithm 
(see Ref.~ \cite{Manna92} and references therein). 
We have considered sizes ranging from $L=32$ to $L=512$ and 
statistics based on a number of configurations 
ranging from $500$ to $100$ respectively.
We have then performed a finite size analysis described in the next subsection
to extract the most reliable values of the exponents.
Our best result for the exponents are $\tau=1.378 \pm 0.002$,
$\psi=1.596 \pm 0.003$, $h=0.633 \pm 0.003$, 
$d_l=1.25 \pm 0.01$, $\phi=2.00 \pm 0.01$ in
excellent agreement with the exact results $\tau=1.375$, $\psi=1.6$,
$h=0.625$, $d_l=1.25$ \cite{Manna92}. This also provides a good test
on the quality of our data analysis. 
Note that the exact value of $\phi$, albeit not known, 
can be derived from the knowledge of $h$ using Eq.~(\ref{cr4}).
This predicts $\phi=2.0$ in perfect agreement with our computed
value. We also note that the obtained results are nearly identical to the
one predicted and numerically generated for a single outlet \cite{Manna96}.

A ST is a self-similar network in that it is undirected
and isotropic. A  quite different choice (directed and anisotropic and hence
self-affine) is a SN. This was again proposed as a
topological model for river networks (see e.g. \cite{Iturbe97})
and the $\tau$ exponent has been exactly determined via  a
mapping to a one-dimensional model for mass aggregation 
\cite{Takayasu91}. A similar mapping to a diffusion-reaction model
also provides the solution in general dimensionality \cite{Swift97}.
Our best results for the critical exponents 
with the same statistics as above are $\tau=1.337 \pm 0.003$,
$\psi=1.49 \pm 0.02$, $h=0.69 \pm 0.02$, $d_l=0.96 \pm 0.01$,
$\phi=1.53 \pm 0.02$, which are in excellent agreement with the
expected values ($\tau$ being exact, the others 
determined via Eq.~(\ref{cr4})),
that is $\tau=4/3$, $\psi=1.5$, $h=0.75$, $d_l=1$, $\phi=1.5$.

\begin{figure}
\centerline{
\epsfxsize=3.5truein
\epsfysize=3.5truein
\epsffile{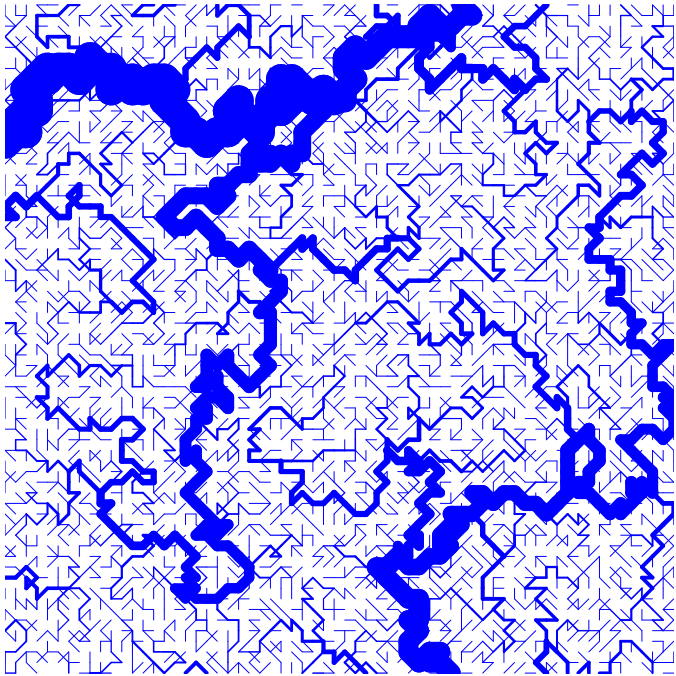}
}
\caption{A spanning tree on a $64 \times 64$ square lattice with
$8$ nearest-neighbors, 4 in the Nord-South and East-West direction and
4 along the diagonal directions. The outlet
is on the West side (outflowing), on the East side there
is an infinite wall, whereas there are periodic boundary conditions
in the North-South direction. The thickness of the line at each point is
proportional to the flow through that point. The seeming loops are
just an artifact of the drawing.
}
\label{fig1}
\end{figure}

\begin{figure}
\centerline{
\epsfxsize=3.5truein
\epsfysize=3.5truein
\epsffile{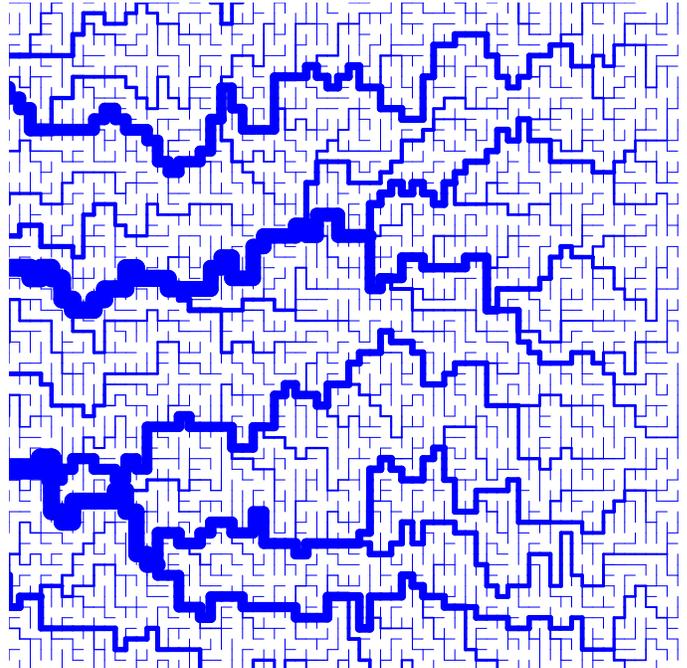}
}
\caption{A Scheidegger network with $L=64$. Boundary conditions are
the same as in Fig.\ref{fig1}. The directness of the network provides
a privileged East-West direction.
}
\label{fig2}
\end{figure}

In Figs.\ref{fig1} and \ref{fig2}, we depict typical patterns for 
a ST and a SN respectively. The presence of multiple outlets
magnifies the main difference between ST and SN. A ST is constructed
in such a way that total freedom is given to the meandering
of streams, the only constraint being that they all terminate
on the same line (West side in the Figures). Typically this
allows the formation of a main big river of size considerably
larger with respect to the others. For SNs, the East-West preferred
direction prevents the forming of such big river and many
smaller rivers are usually present as shown in Fig.\ref{fig2}.

\subsection{The Dynamical Equation: A Self-Consistent solution}
\label{subsec:RDEQ}
As we discussed in Sec. \ref{sec:DEQ}, we seek the stationary states
of the following simplified equation:
\begin{eqnarray}
\frac{\partial h(x,t)}{\partial t}&=& u-Q^{\alpha}(x,t) 
\Bigl[\nabla h(x,t) \Bigr]^2 +\eta(x,t)
\label{res1}
\end{eqnarray}
where $\alpha=2(1-\gamma)$. The stationary  averaged states of Eq. (\ref{res1})
are expected to conform to  Eq. (\ref{sc2}). Despite the apparent simplicity
of the equation, an explicit numerical solution  proves to
be rather slow \cite{Banavar97}. 
The reason for this can be traced back to the particular form
of the erosion term. According to Eq.~(\ref{res1}) only sites
with a non-negligible combination $Q^{\alpha}(x,t) 
\bigl[\nabla h(x,t) \bigr]^2$ will affect the change of the
pattern. As it turns out, this yields a long time transient
during which the elevation profile evolves very slowly.

Since we are mainly interested in stationary states, 
there is a way out from this situation \cite{Banavar97}.
The main idea is to start with an arbitrary network (e.g. a ST or a SN) and
recursively construct the heights starting from the outlets 
with the aid of Eq. (\ref{sc2}).
From the derived landscape, a new network (in  general different
from the original one)  can then be obtained by assuming that at each
site the outward direction is along the steepest descent path
and using Eq.(\ref{int1}). The noise term in Eq. (\ref{res1}) is mimicked 
by the unity term in Eq. (\ref{int1}).
The procedure can then be iterated
until self-consistency is finally achieved. The final configuration is,
by definition, a stationary state of Eq. (\ref{res1}).

\begin{figure}
\centerline{
\epsfxsize=3.5truein
\epsfysize=3.5truein
\epsffile{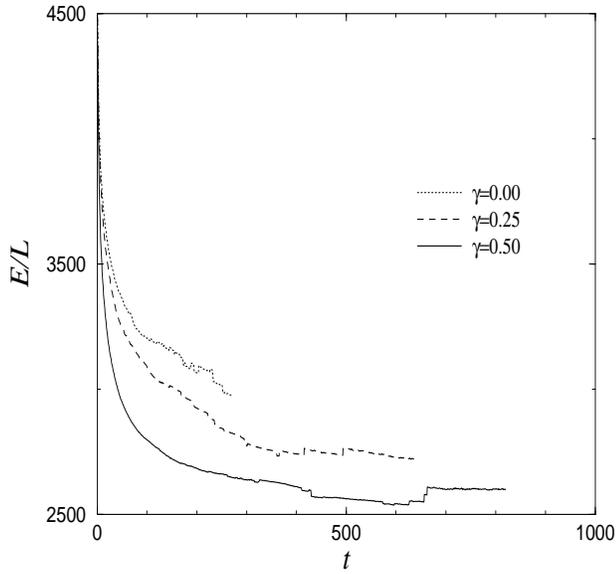}
}
\caption{
Dissipated energy per unit of length, 
as a function of the number of iterations (time $t$) of the self-consistent
solution of the dynamical equation, on starting with the spanning tree of 
Fig~\ref{fig1}. The parameter $\gamma$ is
the value appearing in Eq. (2). The system size 
is $L=512$ and each point of the curve is an average
over all configurations which have gone at least that far in
the number of iterations. All quantities reported in this and 
following figures are dimensionless.
}
\label{fig3}
\end{figure}

The convergence of the above procedure as a function of the number
of iterations is reported in Fig.~\ref{fig3} for various values of
$\gamma$. Two remarks are in order. First we note that the
dissipated energy as obtained from the above SC procedure 
does {\it not} have any physical meaning {\it in the transient state}.
In other words the definition of ``time'' for the independent
variable appearing in Fig.~\ref{fig3} should be considered only as a short-hand
notation for ``number of iterations''. Second it is apparent a
the value $\gamma=0.5$ seems to be the one with the slowest convergence
ratio. This is probably due to the particular role played by the value
$\gamma=0.5$ as it will become clear shortly. 

\begin{figure}
\centerline{
\epsfxsize=3.5truein
\epsfysize=3.5truein
\epsffile{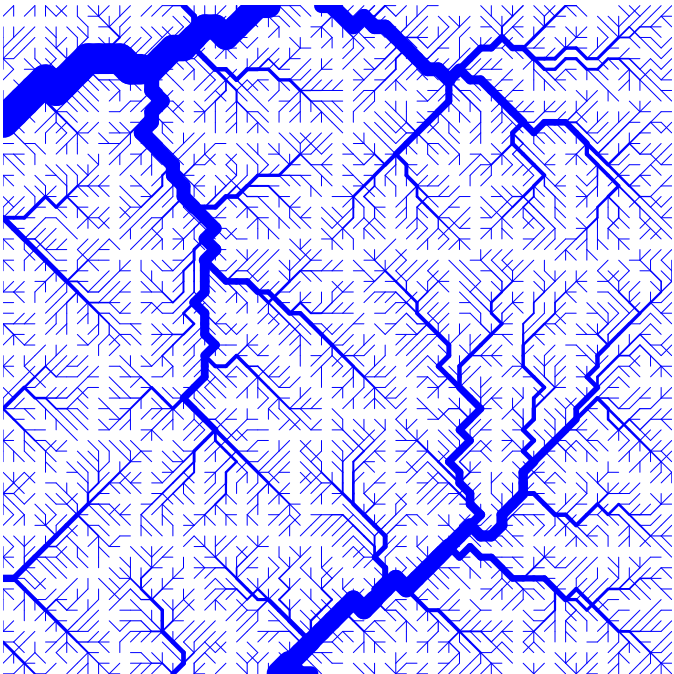}
}
\caption{A typical network obtained as a final output
of the self-consistent procedure described in Sec. 
\ref{subsec:RDEQ}. Here $L=64$, $\gamma=0$, and the initial condition
is the spanning tree of Fig.~\ref{fig1}.
}
\label{fig4}
\end{figure}

\begin{figure}
\centerline{
\epsfxsize=3.5truein
\epsfysize=3.5truein
\epsffile{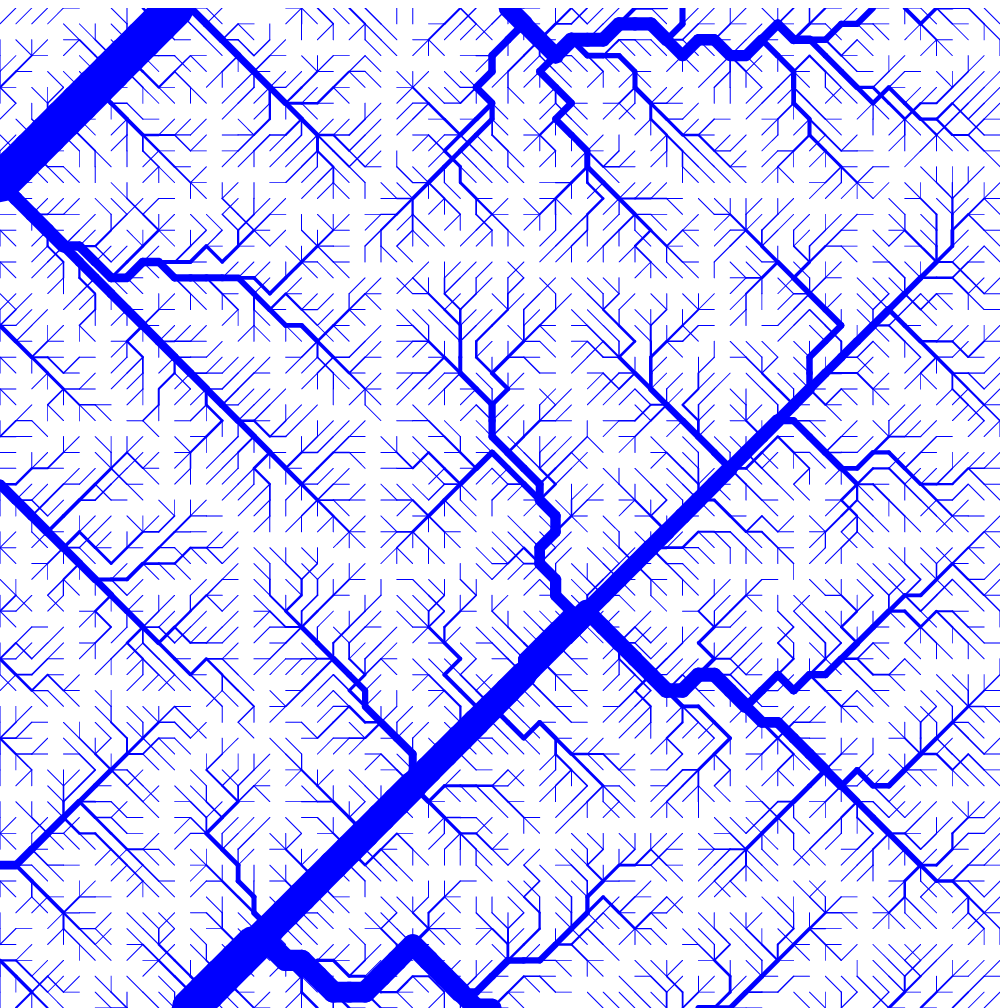}
}
\caption{Same as in Fig.~\ref{fig4}, with $\gamma=0.25$.}
\label{fig5}
\end{figure}

\begin{figure}
\centerline{
\epsfxsize=3.5truein
\epsfysize=3.5truein
\epsffile{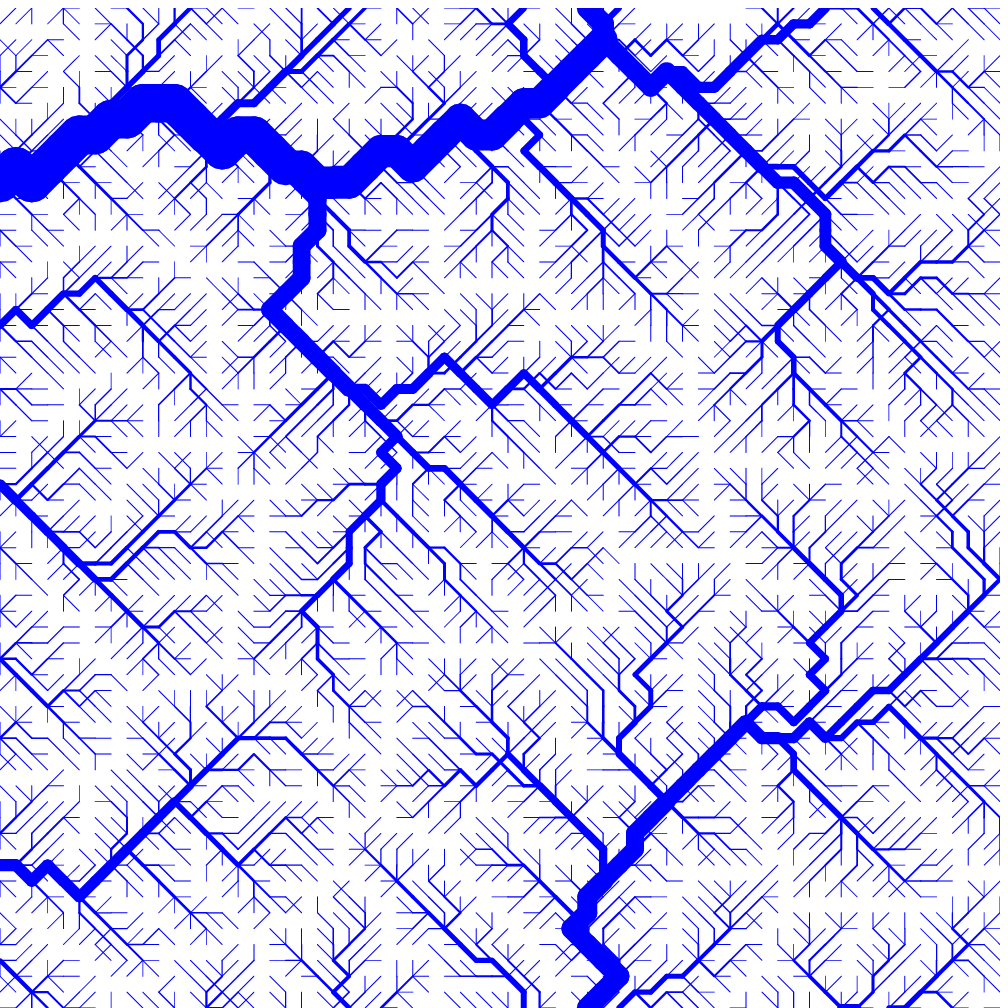}
}
\caption{Same as in Fig.~\ref{fig4}, with $\gamma=0.5$.}
\label{fig6}
\end{figure}

\begin{figure}
\centerline{
\epsfxsize=3.5truein
\epsfysize=3.5truein
\epsffile{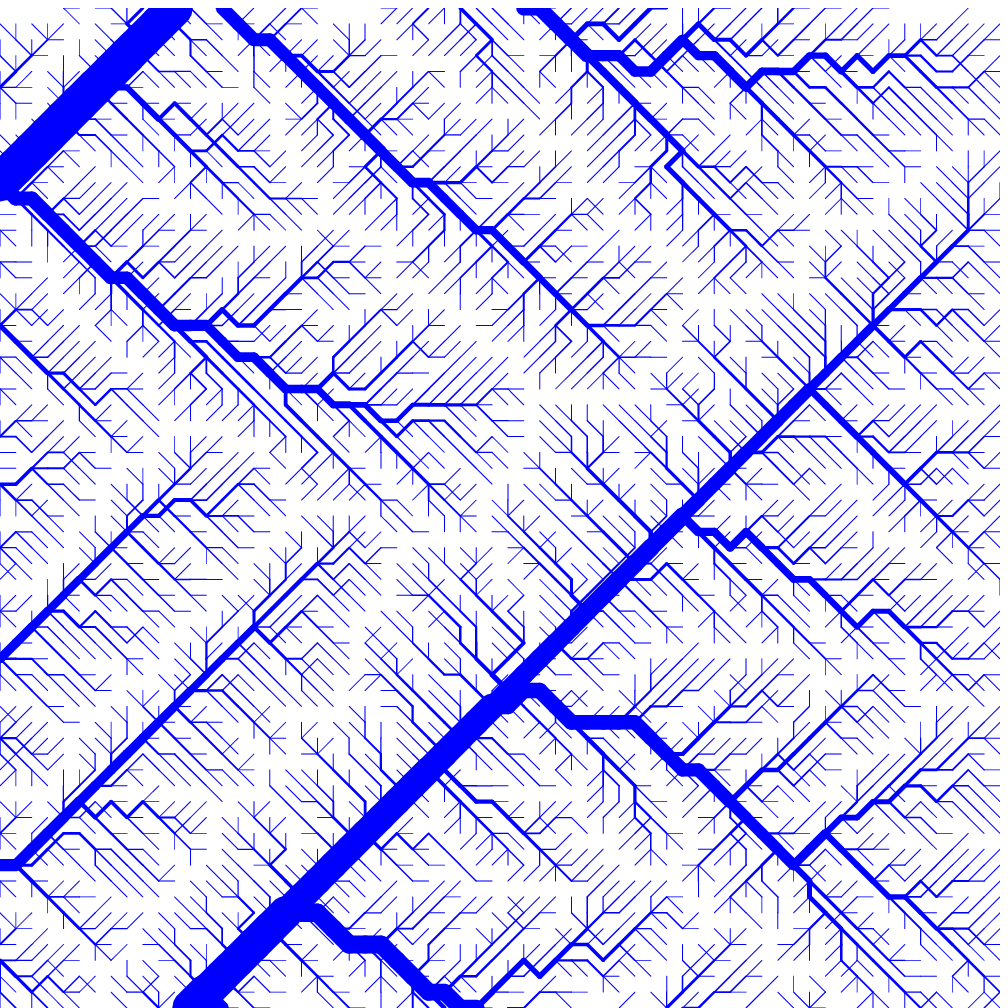}
}
\caption{Same as in Fig.~\ref{fig4}, with $\gamma=0.75$.}
\label{fig7}
\end{figure}

\begin{figure}
\centerline{
\epsfxsize=3.5truein
\epsfysize=3.5truein
\epsffile{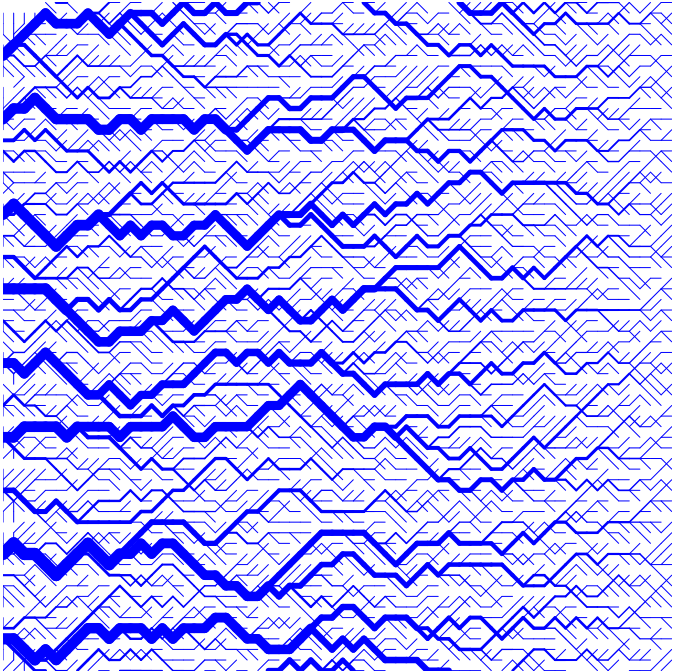}
}
\caption{Same as in Fig.~\ref{fig4}, with $\gamma=1.00$.}
\label{fig8}
\end{figure}

Figs.\ref{fig4}-\ref{fig8} depict typical patterns obtained 
on changing the parameter $\gamma$ in the interval $[0,1]$. The
effect of the value of $\gamma$ on the aggregation pattern is evident.
As $\gamma$ increases from $0$ to $1$, single big rivers draining
the entire basin in a snake-like form are less and less favored.
The pattern for $\gamma=0$ has a strong memory of the original initial tree.
The overall effect of the SC procedure is to disfavor long
meandering of the streams thus providing a self-affine
character to the final tree. The main river becomes rather
straight for $\gamma=0.25$ and the whole pattern appears to
be more symmetric with respect to the case of $\gamma=0$.
A noteworthy feature is that this tendency is inverted as 
$\gamma \to 0.5$ and returns back for $\gamma=0.75$. 
As $\gamma \to 1$ rivers become very directed as one would expect
on the ground that this limit corresponds to the Scheidegger network
\cite{Colaiori97}. The fact that $\gamma=0.5$ most closely resembles
real rivers is a reflection of the natural selection by the
erosional processes of this value of $\gamma$ in terms of Eq. (\ref{sc2})
as it has already been well documented in the literature 
(see e.g. \cite{Iturbe97}).

In our numerical estimates of the exponents, for simplicity, we shall restrict
our attention to the half region $[0,0.5]$ . We also note that this
is the region inaccessible to the analytical scheme of Ref. \cite{Maritan96_2}.

For a more accurate evaluation of the critical exponents  $\tau$
and $\psi$ it proves convenient to introduce the integrated
probabilities:
\begin{eqnarray}
P(a,L) &=& \int_{0}^{a}\ da^{\prime} p(a^{\prime},L)=
a^{1-\tau} F(\frac{a}{L^{\phi}})
\label{res2}
\end{eqnarray}
and
\begin{eqnarray}
\Pi(l,L) &=& \int_{0}^{l}\ dl^{\prime} \pi (l^{\prime},L)=
l^{1-\psi} G(\frac{l}{L^{d_l}})
\label{res3}
\end{eqnarray}
where $F(x)$ and $G(x)$ are related to $f(x)$ and $g(x)$ defined
in Eqs. (\ref{cr2}) and (\ref{cr3}), in
an obvious way. An  efficient way of computing the exponents
is through the so-called ``effective'' 
(sometime also referred to as ``running'') exponents.
In the present case they are defined  as:
\begin{eqnarray}
\tau(a) &=& 1-\frac{\partial\;\log P(a,L)}{\partial \; a}
\label{res4}
\end{eqnarray}
and
\begin{eqnarray}
\psi(l) &=& 1-\frac{\partial\;\log \Pi(l,L)}{\partial \; l}.
\label{res5}
\end{eqnarray}
One then obtains an effective exponent for each value of the
independent variable ($a$ or $l$).

\begin{figure}
\centerline{
\epsfxsize=3.5truein
\epsfysize=3.5truein
\epsffile{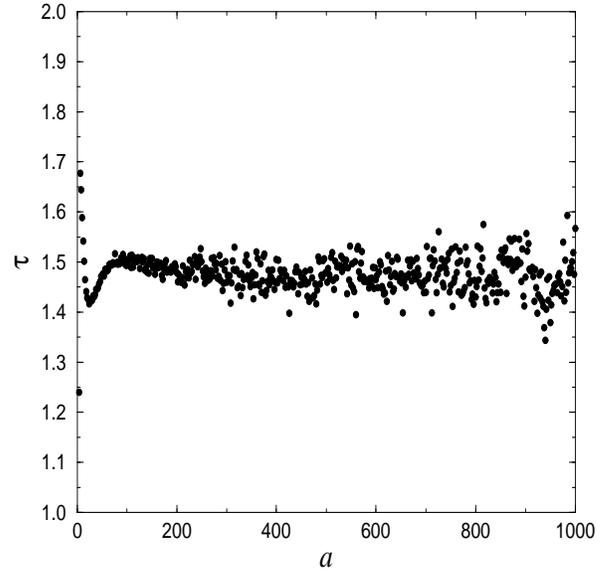}
}
\caption{Effective exponent $\tau$ as a function of the area $a$
in the stationary state of the dynamical equation with $\gamma=0.5$.
Here $L=256$ and the initial condition is 
the spanning tree of Fig.~\ref{fig1}.}
\label{fig9}
\end{figure}

Fig.~\ref{fig9} shows one typical result on a 
$256 \times 256$ lattice.
We can divide the obtained values roughly
in four regions. The first region ($1 \le a < 10$) corresponds to the
region of no scaling. Small rivers belonging to the second region 
($10 \le a < 100$) have an exponent close to the absolute minimum value
$\tau=1.5$. This is consistent with the picture of 
typical rivers
(see Fig.~\ref{fig6}) where small rivers display a marked straightness
similar to the one of the absolute minimum \cite{Colaiori97} and it
means that they quickly assume configurations consistent with the
absolute minimum (independent of the initial conditions). Larger
areas are associated with larger rivers which have longer memory of the initial
condition (ST in the present case). Hence the corresponding
exponent is sensibly smaller (closer to the ST value $1.38$). The last
region corresponds to the finite size cut-off and must be discarded.

After discarding the first and forth regions 
the obtained values can then be grouped into local bins  and
a local average exponent can be associated with each of them.
Statistical fluctuations within each box then yield
an estimate of error bars.
This provides our best estimate of the exponent for each
value of $L$ and a simple $1/L$ extrapolation is then carried out
to extract the final values. This is depicted in Fig.~\ref{fig10} 
and the corresponding
best estimates of this method are reported in Table~\ref{table1}. 
An analogous procedure
leads to the best estimates for $\psi$ as reported again in Table~\ref{table1}.
Sizes and statistics are identical to those considered for the initial
conditions and hence these simulations are rather time consuming.

One can notice a weak dependence on $\gamma$ for both $\tau$ and $\psi$.
One way of computing $\phi$  and  $d_l$ is through the collapse plots
of the probabilities $P(a,L)$ and $\Pi(l,L)$. 
However we have found that a satisfactory collapse can achieve only within
a limited range of the appropriate variable ($a/L^{\phi}$ and $l/L^{d_l}$).
Hence we have opted for an alternative scheme hinging on the
calculation of the following ratios:
\begin{eqnarray}
M^{q}_a(L)&\equiv&\frac{ \left \langle a^{q+1} \right \rangle_a}
{\left \langle a^{q} \right \rangle_a} \sim L^{\phi} \;\;\; q=1,2,...
\label{res6}
\end{eqnarray}
where averages are over the probability densities $p(a,L)$ 
(of the maximum river) and
the $L$ dependence is straightforward \cite{Maritan96_1}.
A similar relationship holds for the lengths:
\begin{eqnarray}
M^{q}_l(L)&\equiv&\frac{ \left \langle l^{q+1} \right \rangle_l}
{\left \langle l^{q} \right \rangle_l} \sim L^{d_{l}} \;\;\; q=1,2,...
\label{res7}
\end{eqnarray}
The results for $q=1,2,3$ are reported in Table \ref{table3} for the
exponent $\phi$ and in Table \ref{table4} for the exponent $d_l$.
Surprisingly the values for $\phi$ are consistently larger that
the space-filling value $\phi=2$ which is expected to be the upper bound
for this exponent. This is probably due to a deficiency of this procedure
during a cross-over from a self-similar regime (the initial ST) to
a self-affine pattern (the final configuration). Indeed we shall see
later on that this feature is not present when one starts with a self-affine
network (e.g. a SN) from the outset. A seemingly large value of $d_l$
is appearing probably due to the same reason. Overall, one
can notice a rather weak (if any) dependence on $\gamma$ and almost
no dependence on $q$.

Finally Hack's exponent $h$ was computed from its definition Eq.~(\ref{cr1})
using an effective exponent method to be described below.

Let us assume a power-law dependence for a generic function as given by
\begin{eqnarray}
f(x)&\sim& x^{\theta} \qquad \qquad \theta > 0
\label{res8}
\end{eqnarray}
On integrating $f(x)$ (between  lower and upper limits, say $x_0$ and
$x$), we find
\begin{eqnarray}
\theta&=& \frac{\Bigl[x f(x)-x_0 f(x_0) \Bigr]}{\int_{x_0}^{x} dz f(z)} -1.
\label{res9}
\end{eqnarray} 
The effective exponent obtained using Eq.(\ref{res9}) in Eq.(\ref{cr1})
with $f(x)\equiv a $, $x \equiv l$, and
$\theta \equiv 1/h$, can then be analyzed with the same procedure (local
average plus $1/L$ extrapolation) outlined before. We note that
for $a$ we have used an averaged value over all areas corresponding to the
same length.
The final results are reported in Table~\ref{table1} and one can see that
there is an overall good agreement with scaling laws.

\subsection{OCN}
\label{subsec:ROCN}
The minimization of the energy functional Eq.~(\ref{ocn2}) goes through
an algorithm akin to the one exploited by Sun et. al \cite{Sun94}. It 
is based on the following steps:
\begin{description}
\item{1)} An initial configuration (ST or SN) is generated and its
dissipated energy is computed according to Eq.~(\ref{ocn2}). By definition,
this initial network has no loops.
\item{2)} A link is randomly selected and its local outflow is also randomly
chosen.
\item{3)} This new configuration is tested for loop creation. If a loop
has been created, the configuration is rejected and step 2) is repeated.
\item{4)} The energy of the new candidate configuration is computed. If
it is smaller than the previous one, the new configuration is accepted,
otherwise it is rejected.
\item{5)} Steps 2)-4) are repeated until the energy does not change 
within a given tolerance.
\end{description}
The final configuration is regarded as a local minimum.
This scheme is patterned after 
a standard Metropolis algorithm at zero temperature, and
averages are over many different configurations
(ranging from $500$ at $L=32$ to $100$ at $L=256$). 
Fig.~\ref{fig11} depicts the dissipated energy per unit of length
as a function of the convergence ``time'' (i.e. the number of total iteration
of the algorithm). The similarity with patterns obtained from the
SC procedure is evident. Here, however, the typical convergence time is much
longer than before.

Critical exponents
are computed with the same prescription given in the
previous subsection. In Fig.~\ref{fig12}
we show the resulting $1/L$ extrapolation. The final best estimates
are reported in Table~\ref{table1}.

Once more, a weak dependence on $\gamma$
($\tau$ increases as $\gamma$ increases) can be noticed. 
All other exponents
are consistent with scaling relations Eq. (\ref{cr4}). 
It is worth stressing that
exponents are nearly consistent (at the edge of statistical errors) with those 
previously obtained from the dynamical equation.

Regarding the exponents $\phi$ and $d_l$, they can be found in
Tables \ref{table3} and \ref{table4} respectively. Here too
the same feature, discussed in \ref{subsec:RDEQ} in connection
with the exponent values of $\phi$ and $d_l$, applies.

\subsection{Independence of the initial condition}
\label{subsec:independence}
Our final task is a test of the sensitivity of critical exponents to
the initial conditions. To this aim we have changed the
initial condition from a ST to a SN for both the dynamical equation
and the OCN. The main difference between the
two initial conditions is that while the latter is a directed network, 
the former is not. Table~\ref{table2} reports the comparison 
for the case $\gamma=0.5$, and Fig.~\ref{fig13} depicts the network
resulting from the SC scheme, for
a typical final state. 
Despite the obvious memory of the initial network (see Fig.~\ref{fig2}), 
the typical distance
among big rivers is larger than the initial one. This in turn yields
a higher value for $H$ and hence non-trivial exponents.

It is remarkable that although
these patterns appear to be considerably different from those
obtained when starting with a ST, the two sets of critical exponents
are nevertheless consistent within the statistical errors.

As a final comment, we find in this case values of $\phi$ and $d_l$ 
in very good agreement
with scaling predictions. 
The results for different ratios $q$ in the case $\gamma=0.5$ 
for both the SC and the OCN, along with the comparison with
the corresponding values stemming from STs are reported in Tables
\ref{table5} and \ref{table6} respectively.

\section{Conclusions}
\label{sec:conclusions}
In this paper we have addressed the issue of the existence and robustness
of the universality class associated with landscapes corresponding
to local minimal energies.

To this aim we have first extended previous studies for both the SC solution
of a Langevin equation and the OCNs variational methods. Higher sizes
and statistics have been exploited in both cases and, (to our knowledge)
for the first time, 
the most physical procedure of basing the statistics only on the river with 
largest flow, has been implemented.

Second, we have monitored the dependence of critical exponents on
a parameter $\gamma$ associated with the slope-area law in one case (SC)
and with the dissipated energy in the other (OCN).
Our results give compatible critical exponents between SC and OCN
within the error bars, but a weak and similar dependence on $\gamma$
appears in both models.
 
Finally we have tested the stability of the obtained results
for both the SC and the OCN with respect to changes in the 
initial conditions. Although the obtained final patterns display a dependence 
on initial conditions, critical exponents appear to be insensitive
to this dependence.

As a by-product of our investigation, we have found that the SC solution
of the dynamical equation is a very powerful method to investigate 
river networks as
it is capable of providing useful informations on the stationary state
in a simple and physical way. Another interesting
point, from a numerical point of view, is that this procedure typically
achieves convergence much faster with respect to OCN scheme.

\begin{figure}
\centerline{
\epsfxsize=3.5truein
\epsfysize=3.5truein
\epsffile{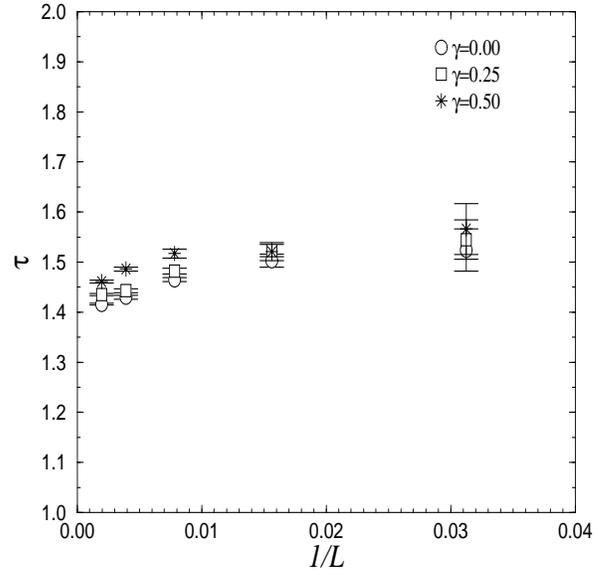}
}
\caption{Finite size $1/L$ extrapolation for the value of $\tau$ obtained
with the dynamical equation  
for the cases $\gamma=0,0.25,0.5$. In all cases the initial
condition is the spanning tree of Fig.~\ref{fig1}.}
\label{fig10}
\end{figure}

\begin{figure}
\centerline{
\epsfxsize=3.5truein
\epsfysize=3.5truein
\epsffile{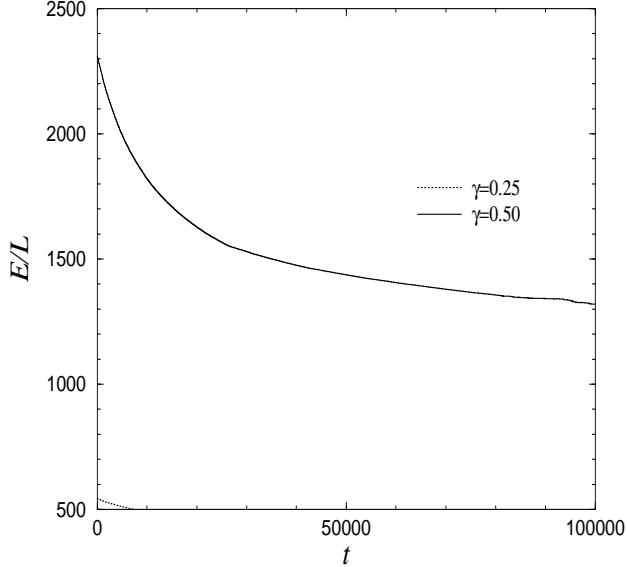}
}
\caption{Dissipated energy per unit of length, 
as a function of the number of iterations in the OCN procedure.
Here the size is $L=256$ and again each point of the curves
are averaged over all configurations which have reached that
time $t$. The initial condition is the spanning tree of Fig.~\ref{fig1}.}
\label{fig11}
\end{figure}

\begin{figure}
\centerline{
\epsfxsize=3.5truein
\epsfysize=3.5truein
\epsffile{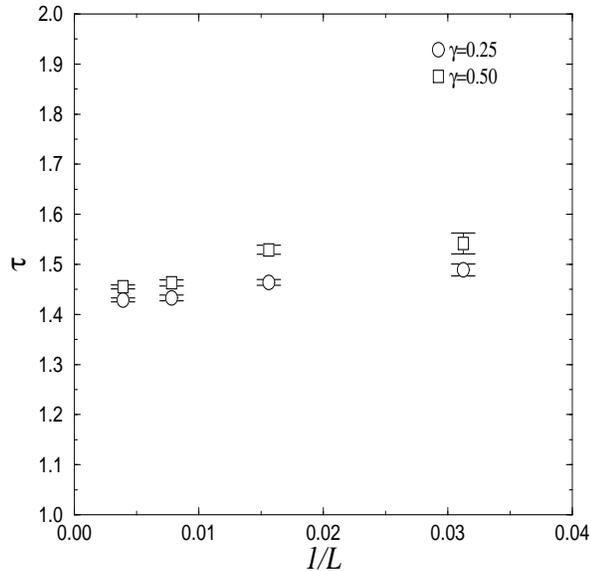}
}
\caption{Finite size $1/L$ extrapolation for the value of $\tau$ obtained
with the OCN ($T=0$) for the cases $\gamma=0.25,0.5$. In both cases the
initial condition is the spanning tree of Fig.~\ref{fig1}.}
\label{fig12}
\end{figure}

\begin{figure}
\centerline{
\epsfxsize=3.5truein
\epsfysize=3.5truein
\epsffile{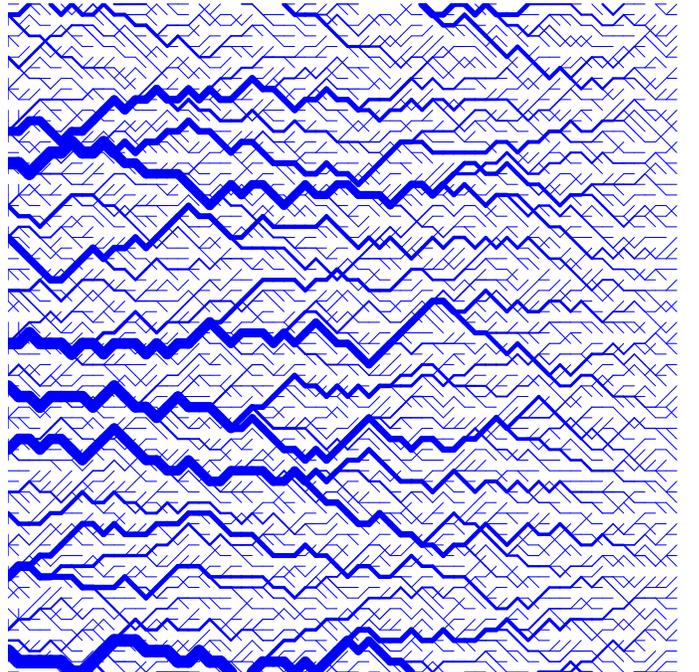}
}
\caption{A typical $64 \times 64$ network obtained as a final output
of the self-consistent procedure described in Sec. 
\protect{\ref{subsec:RDEQ}} upon starting with a the Scheidegger
network of Fig.~\ref{fig2}. 
}
\label{fig13}
\end{figure}

In view of our results we can now summarize the arguments favoring
and disfavoring the appearance of a new universality class associated
with local minima.
As we mentioned in our discussion of Fig.~\ref{fig10} the typical
evolution of a network appears to depend on the
considered length scale. Small rivers very quickly settle to their
final state whereas much longer time is required to large rivers
to forget their initial conditions. This is also reflected
by the difficulty in collapsing the distribution probabilities
of both $a$ and $l$ into a single plot for a reasonably
extended range of the corresponding variable. 
It is then possible that although only {\it two} universality
classes (one associated with the initial condition and the other to
the absolute minimum)
are present, an intermediate universality class sets on due
to both the averaging over different regimes and the difficulty of reaching
the absolute minimum.

On the other hand, in support of the existence of a new universality class, we
cite the fact that critical exponents are different from those of both
ST and SN -- on starting with these initial conditions, critical exponents
in the final configurations
clearly deviate from their initial values. Furthermore all exponents 
are found to be robust and obey to scaling relations summarized 
in Sec.\ref{sec:Exp}.

Overall we believe that the evidence suggested by our results favors
this second possibility which has also been hinted into a different
context \cite{Tadic98}. In this respect the weak $\gamma$ dependence of
critical exponents remains unexplained.

It would be interesting to generalize the present calculation
in two aspects. For the dynamical equation, 
it would be instructive to tackle the problem of the
explicit solution of Eq. (\ref{sc1}). This route has the advantage that
Eq. (\ref{sc2}) (the key relation between erosional process
and network topology) can be then {\it derived} rather than {\it assumed}
as in the self-consistent procedure. Similarly, a parallel calculation 
could also be
implemented in the OCN framework, upon starting with the more
general expression given in Eq. (\ref{ocn1}).
 
\acknowledgements
It is a pleasure to thank Jayanth Banavar, Amos Maritan and Andrea Rinaldo for
many enlighting discussions, innumerable suggestions and a critical
reading of the manuscript. Financial support from the Italian MURST (Ministero
dell'Universit\'a e della Ricerca Scientifica e Tecnologica) and INFM (Istituto
Nazionale di Fisica della Materia) is gratefully acknowledged.
 


\begin{table}
\caption{Critical exponents $\tau$, $\psi$ and $h$ as a function
of $\gamma$ for both the Self-Consistent (SC) procedure 
and the optimal channel networks (OCN) at zero temperature.
The values in the parenthesis are those obtained from the
computed value of $\tau$ and using the scaling relations.
For $\gamma=0$, the OCN exponents 
reported are the exact ones corresponding to the initial conditions
(a spanning tree in the present case) since the energy is a constant.}

\begin{tabular}{lcccccc}
\multicolumn{1}{l}{$\gamma$}&
\multicolumn{1}{c}{$\tau_{OCN}$}&
\multicolumn{1}{c}{$\tau_{SC}$}&
\multicolumn{1}{c}{$\psi_{OCN}$}&
\multicolumn{1}{c}{$\psi_{SC}$}&
\multicolumn{1}{c}{$h_{OCN}$}&
\multicolumn{1}{c}{$h_{SC}$}\\
\hline
0.00 & 1.38 & $1.40 \pm 0.01$ & 1.6 &
$1.69 \pm 0.03$ ($1.67$) & 0.625  & $0.61 \pm 0.02$ ($0.60$)  \\
0.25 & $1.42 \pm 0.01$ & $1.42 \pm 0.01$ & $1.71 \pm 0.07$ ($1.72$) &
$1.68 \pm 0.05$ ($1.72$) & $0.64 \pm 0.03$ ($0.58$) & $0.58 \pm 0.01$ ($0.58$)  \\
0.50 & $1.44 \pm 0.01$ & $1.46 \pm 0.01$ & $1.8 \pm 0.1$ ($1.79$) &
$1.82 \pm 0.05$ ($1.85$) & $0.61 \pm 0.03$ ($0.56$)  & $0.56 \pm 0.01$ ($0.54$) \\
\hline
\end{tabular}
\label{table1}
\end{table}
\begin{table}
\caption{Critical exponents $\tau$, $\psi$ and $h$ for $\gamma=0.5$
as a function of the initial conditions. As in the text, ST and SN stand
for Spanning Trees and Scheidegger Network respectively. SC and OCN
have the same meaning as above.
Values in parenthesis are scaling predictions.
}
\begin{tabular}{lcccccc}
\multicolumn{1}{l}{}&
\multicolumn{1}{c}{$\tau_{OCN}$}&
\multicolumn{1}{c}{$\tau_{SC}$}&
\multicolumn{1}{c}{$\psi_{OCN}$}&
\multicolumn{1}{c}{$\psi_{SC}$}&
\multicolumn{1}{c}{$h_{OCN}$}&
\multicolumn{1}{c}{$h_{SC}$}\\
\hline
\text{ST} & $1.44 \pm 0.01$& $1.46 \pm 0.01$ & $1.8 \pm 0.1$
($1.79$) & $1.82 \pm 0.05$ ($1.85$)& $0.61 \pm 0.03$ ($0.56$) &
 $0.56 \pm 0.01$ ($0.54$)  \\
\text{SN} &$1.44 \pm 0.03$ & $1.43 \pm 0.02$ & $1.8 \pm 0.2$ 
($1.79$)& $1.7 \pm 0.1$ ($1.75$) & $0.61 \pm 0.03$ ($0.56$) & 
$0.55 \pm 0.02$ ($0.57$) \\
\hline
\end{tabular}
\label{table2}
\end{table}
\begin{table}
\caption{Critical exponent $\phi$ stemming from both the
SC procedure and the OCN scheme, 
as a function of the value of $\gamma$
and of the value of the order $q$ of ratios defined in
Eqs. (\ref{res4}) and (\ref{res5}). 
}
\begin{tabular}{lcccccc}
\multicolumn{1}{l}{$q~$}&
\multicolumn{1}{c}{$\gamma=0.00 ~ (\mbox{SC })$}&
\multicolumn{1}{c}{$\gamma=0.25 ~ (\mbox{SC })$}&
\multicolumn{1}{c}{$\gamma=0.50 ~ (\mbox{SC })$}&
\multicolumn{1}{c}{$\gamma=0.00 ~ (\mbox{OCN})$}&
\multicolumn{1}{c}{$\gamma=0.25 ~ (\mbox{OCN})$}&
\multicolumn{1}{c}{$\gamma=0.50 ~ (\mbox{OCN})$}\\
\hline
$1$ & $2.1 \pm 0.1$& $2.2 \pm 0.1$ & $2.2 \pm 0.1$
& - & $2.0 \pm 0.1$ & $2.1 \pm 0.1$   \\
$2$ & $2.1 \pm 0.1$& $2.2 \pm 0.1$ & $2.2 \pm 0.1$
& - & $2.0 \pm 0.1$ & $2.1 \pm 0.1$   \\
$3$ & $2.1 \pm 0.1$& $2.2 \pm 0.1$ & $2.3 \pm 0.1$
& - & $2.0 \pm 0.1$ & $2.0 \pm 0.1$   \\
\hline
\end{tabular}
\label{table3}
\end{table}
\begin{table}
\caption{Critical exponent $d_l$ obtained from both the
SC procedure and the OCN scheme, 
as a function of the value of $\gamma$
and of the value of the order $q$ of ratios (\ref{res4}) and (\ref{res5}) . 
}
\begin{tabular}{lcccccc}
\multicolumn{1}{l}{$q~$}&
\multicolumn{1}{c}{$\gamma=0.00 ~ (\mbox{SC })$}&
\multicolumn{1}{c}{$\gamma=0.25 ~ (\mbox{SC })$}&
\multicolumn{1}{c}{$\gamma=0.50 ~ (\mbox{SC })$}&
\multicolumn{1}{c}{$\gamma=0.00 ~ (\mbox{OCN})$}&
\multicolumn{1}{c}{$\gamma=0.25 ~ (\mbox{OCN})$}&
\multicolumn{1}{c}{$\gamma=0.50 ~ (\mbox{OCN})$}\\
\hline
$1$ & $1.3 \pm 0.1$& $1.3 \pm 0.1$ & $1.1 \pm 0.1$
& - & $1.2 \pm 0.1$ & $1.3 \pm 0.1$   \\
$2$ & $1.3 \pm 0.1$& $1.4 \pm 0.1$ & $1.2 \pm 0.1$
& - & $1.3 \pm 0.1$ & $1.3 \pm 0.1$   \\
$3$ & $1.3 \pm 0.1$& $1.4 \pm 0.1$ & $1.2 \pm 0.1$
& - & $1.3 \pm 0.1$ & $1.3 \pm 0.1$   \\
\hline
\end{tabular}
\label{table4}
\end{table}
\begin{table}
\caption{Comparison between exponents $\phi$  as obtained from both the
SC procedure and the OCN scheme, 
as a function of the value of the initial conditions 
and of the value of the order $q$ of ratios (\ref{res4}) and (\ref{res5}).
As in the text, ST stands for Spanning Trees and SN for Scheidegger Network.
Here $\gamma=0.5$. 
}
\begin{tabular}{lcccc}
\multicolumn{1}{l}{$q~$}&
\multicolumn{1}{c}{$\phi_{\mathrm{ST}} ~ (\mbox{SC })$}&
\multicolumn{1}{c}{$\phi_{\mathrm{SN}} ~ (\mbox{SC })$}&
\multicolumn{1}{c}{$\phi_{\mathrm{ST}} ~ (\mbox{OCN})$}&
\multicolumn{1}{c}{$\phi_{\mathrm{SN}} ~ (\mbox{OCN})$} \\
\hline
$1$ & $2.2 \pm 0.1$& $1.8 \pm 0.1$ & $2.1 \pm 0.1$ & $1.7 \pm 0.1$   \\
$2$ & $2.2 \pm 0.1$& $1.8 \pm 0.1$ & $2.1 \pm 0.1$ & $1.7 \pm 0.1$   \\
$3$ & $2.3 \pm 0.1$& $1.8 \pm 0.1$ & $2.0 \pm 0.1$ & $1.7 \pm 0.1$   \\
\hline
\end{tabular}
\label{table5}
\end{table}
\begin{table}
\caption{Comparison between exponents $d_l$  resulting from both the
SC procedure and the OCN scheme, 
as a function of the value of the initial conditions 
and of the value of the order $q$ of ratios (\ref{res4}) and (\ref{res5}).
As in the text, ST stands for Spanning Trees and SN for Scheidegger Network.
Here $\gamma=0.5$. 
}
\begin{tabular}{lcccc}
\multicolumn{1}{l}{$q~$}&
\multicolumn{1}{c}{${d_l}_{\mathrm{ST}} ~ (\mbox{SC })$}&
\multicolumn{1}{c}{${d_l}_{\mathrm{SN}} ~ (\mbox{SC })$}&
\multicolumn{1}{c}{${d_l}_{\mathrm{ST}} ~ (\mbox{OCN})$}&
\multicolumn{1}{c}{${d_l}_{\mathrm{SN}} ~ (\mbox{OCN})$} \\
\hline
$1$ & $1.1 \pm 0.1$& $0.9 \pm 0.1$ & $1.3 \pm 0.1$ & $0.9 \pm 0.1$   \\
$2$ & $1.2 \pm 0.1$& $1.0 \pm 0.1$ & $1.3 \pm 0.1$ & $1.0 \pm 0.1$   \\
$3$ & $1.2 \pm 0.1$& $1.0 \pm 0.1$ & $1.3 \pm 0.1$ & $1.0 \pm 0.1$   \\
\hline
\end{tabular}
\label{table6}
\end{table}




\begin{references}
\bibitem{Iturbe97} I. Rodrigues-Iturbe and A. Rinaldo, {\it Fractal
River Basins: Chance and Self-Organization} (Cambridge University
Press, Great Britain, 1997).
\bibitem{Rinaldo93} A. Rinaldo, I. Rodrigues-Iturbe, R. Rigon and R.L. Bras,
Phys. Rev. Lett. {\bf 70}, 822 (1993).
\bibitem{Meakin91} P. Meakin, J. Feder, and T. Jossang, Physica A 
{\bf 176}, 409 (1991).
\bibitem{Maritan96_1} A. Maritan, A. Rinaldo, R. Rigon, A. Giacometti
and I. Rodrigues-Iturbe, Phys. Rev. E {\bf 53}, 1510 (1996).
\bibitem{Banavar97} J.R. Banavar, F. Colaiori, A. Flammini, A. Giacometti,
A. Maritan and A. Rinaldo, Phys. Rev. Lett. {\bf 78}, 4522 (1997).
\bibitem{Sinclair96} K. Sinclair and R. C. Ball, Phys. Rev. Lett. 
{\bf 76}, 3360 (1996).
\bibitem{Somafai97} E. Somafai and L. M. Sander, Phys. Rev. E {\bf 56},
R5 (1997).
\bibitem{Dodds99} P. S. Dodds and D. H. Rothman, Phys. Rev. E {\bf 59},
4865 (1999).
\bibitem{Leopold62} L.B.Leopold and W.B.Langbein, U.S. Geol. Surv. Prof.
{bf 500-A},1 (1962).
\bibitem{Marsili96} M. Marsili, A. Maritan, F. Toigo and J. R. Banavar,
Rev. Mod. Phys. {\bf 68}, 963 (1996).
\bibitem{Pastor98} R. Pastor-Satorras and D. H. Rothman, Phys. Rev. Lett.
{\bf 19}, 4349 (1998).
\bibitem{Metiver99} F. M\'etiver, Phys. Rev. E {\bf 60}, 5827 (1999)
\bibitem{Giacometti95} A. Giacometti, A. Maritan, and J. R. Banavar,
Phys. Rev. Lett. {\bf 75}, 577 (1995).
\bibitem{Iturbe92} I. Rodriguez-Iturbe, A. Rinaldo, R. Rigon, R. L. Bras, and
E. Ijjasz-Vasquez, Water Resour. Res. {\bf 28}, 1095 (1992);
I. Rodriguez-Iturbe, A. Rinaldo, R. Rigon, R. L. Bras, and E. Ijjasz-Vasquez,
Geophys. Res. Lett. {\bf 19}, 889 (1992); A. Rinaldo et. al, Water Resour. Res.
{\bf 28}, 2183 (1992).
\bibitem{Kirkpatrick83} S. Kirkpatrick, C.D. Gelatt, and M. P. Vecchi,
Science {\bf 220}, 671 (1983).
\bibitem{Sun94} T. Sun, P. Meakin and T. Jossang, Phys. Rev. E {\bf 49},
4865 (1994).
\bibitem{Maritan96_2} A. Maritan, F. Colaiori, A. Flammini, M. Cieplak and
J. R. Banavar, Science {\bf 272}, 984 (1996).
\bibitem{Colaiori97} F. Colaiori, A. Flammini, A. Maritan and J. R. Banavar,
Phys. Rev. E {\bf 55}, 1298 (1997).
\bibitem{Mandelbrot83} The Peano basin is discussed e.g. in  B.B. Mandelbrot
{\it The Fractal Geometry of Nature}, (Freeman, New York 1983). Some
topological properties in the river network framework have been addressed in 
A. Marani, R. Rigon and A. Rinaldo, Water Resour.
Res. {\bf 27}, 3041 (1991).
\bibitem{note1} We note that $\gamma=0$ and $\gamma=1$ correspond
to known and exactly solvable models (spanning trees and Scheidegger
network respectively) and that the only physically acceptable values 
of $\gamma$ lie between these two limiting cases.
\bibitem{Hack57} J. T. Hack, U.S. Geological Survey Professional Papers
{\bf 294-B}, 45 (1957).
\bibitem{Cieplak98} M. Cieplak, A. Giacometti, A. Maritan, A. Rinaldo,
I. Rodrigues-Iturbe and J. R. Banavar, J. Stat. Phys. {\bf 91}, 1 (1998).
\bibitem{Caldarelli97} G. Caldarelli, A. Giacometti, A. Maritan, 
I.Rodrigues-Iturbe and A. Rinaldo, Phys. Rev. E {\bf 55}, R4865 (1997).
\bibitem{Wu82} F. Y. Wu, Rev. Mod. Phys. {\bf 54}, 235 (1982).
\bibitem{Manna92} S. S. Manna, D. Dhar and S. N. Majumadar, Phys. Rev. A
{\bf 46}, R4471 (1992).
\bibitem{Manna96} S.S. Manna and B. Subramanian, Phys. Rev. Lett. {\bf 76},
3460 (1996).
\bibitem{Takayasu91} M. Takayasu, H. Takayasu and G. Huber, J. Stat. Phys.
{\bf 65}, 725 (1991).
\bibitem{Swift97} M. R. Swift, F. Colaiori, A. Flammini, A. Maritan, 
A. Giacometti and J. R. Banavar, Phys. Rev. Lett. {\bf 79}, 3278 (1997).
\bibitem{Tadic98} B. Tadic, Phys. Rev. E {\bf 58}, 168 (1998).
\end{references}
\end{document}